\documentclass[sn-mathphys-num]{sn-jnl}%

\usepackage[dvipsnames]{xcolor}
\usepackage{graphicx}%
\usepackage{multirow}%
\usepackage{amsmath,amssymb,amsfonts}%
\usepackage{amsthm}%
\usepackage{mathrsfs}%
\usepackage[title]{appendix}%
\usepackage{xcolor}%
\usepackage{textcomp}%
\usepackage{manyfoot}%
\usepackage{booktabs}%
\usepackage{algorithm}%
\usepackage{algorithmicx}%
\usepackage{algpseudocode}%
\usepackage{listings}%
\usepackage[normalem]{ulem}
\usepackage{color,soul}
\usepackage{tabularx}

\theoremstyle{thmstyleone}%
\theoremstyle{thmstyletwo}%

\theoremstyle{thmstylethree}%

\begin{document}

\title[Article Title]{Simulation of Transcatheter Therapies for Atrioventricular Valve Regurgitation in an Open-Source Finite Element Simulation Framework}

\author[1]{\fnm{Seda} \sur{Aslan}}\email{aslans1@chop.edu}
\equalcont{These authors contributed equally to this work.}
\author[1]{\fnm{Nicolas R.} \sur{Mangine}}\email{manginen@chop.edu}
\equalcont{These authors contributed equally to this work.}
\author[1,2]{\fnm{Devin W.} \sur{Laurence}}\email{laurenced@chop.edu}
\author[1]{\fnm{Patricia M.} \sur{Sabin}}\email{sabinp@chop.edu}

\author[1,3,4]{\fnm{Wensi} \sur{Wu}}\email{wuw4@chop.edu}

\author[1]{\fnm{Christian} \sur{Herz}}\email{herzc@chop.edu}

\author[1,2]{\fnm{Justin S.} \sur{Unger}}\email{ungerj@chop.edu}

\author[5,6]{\fnm{Steve A.} \sur{Maas}}\email{steve.maas@utah.edu}
\author[2]{\fnm{Matthew J.} \sur{Gillespie}}\email{GILLESPIE@chop.edu}

\author[5,6]{\fnm{Jeffrey A.} \sur{Weiss}}\email{jeff.weiss@utah.edu}

\author*[1,2]{\fnm{Matthew A.} \sur{Jolley}}\email{jolleym@chop.edu}

\affil[1]{\orgdiv{Department of Anesthesiology and Critical Care Medicine}, \orgname{Children's Hospital of Philadelphia}, \orgaddress{\city{Philadelphia}, \state{PA}, \country{USA}}}

\affil[2]{\orgdiv{Division of Cardiology}, \orgname{Children's Hospital of Philadelphia}, \orgaddress{\city{Philadelphia}, \state{PA}, \country{USA}}}

\affil[3]{\orgdiv{Department of Mechanical Engineering and Applied Mechanics}, \orgname{University of Pennsylvania}, \orgaddress{\city{Philadelphia}, \state{PA}, \country{USA}}}

\affil[4]{\orgdiv{Cardiovascular Institute}, \orgname{Children's Hospital of Philadelphia}, \orgaddress{\city{Philadelphia}, \state{PA}, \country{USA}}}

\affil[5]{\orgdiv{Department of Biomedical Engineering}, \orgname{University of Utah}, \orgaddress{\city{Salt Lake City}, \state{UT}, \country{USA}}}

\affil[6]{\orgdiv{Scientific Computing Institute}, \orgname{University of Utah}, \orgaddress{\city{Salt Lake City}, \state{UT}, \country{USA}}}

\abstract{\textbf{Purpose:} Transcatheter edge-to-edge repair (TEER) and annuloplasty devices are increasingly used to treat mitral valve regurgitation, yet their mechanical effects and interactions remain poorly understood. This study aimed to establish an open-source finite element modeling (FEM) framework for simulating patient-specific mitral valve repairs and to evaluate how TEER, annuloplasty, and combined strategies influence leaflet coaptation and valve mechanics. A central objective was to demonstrate how such simulations may support surgical planning by identifying optimal interventions.

\textbf{Methods:} A patient-specific mitral valve model was reconstructed using SlicerHeart and 3D Slicer. Four G4 MitraClip geometries were modeled and deployed in FEBio to capture leaflet grasp and subsequent clip–leaflet motion under physiologic pressurization. CardioBand annuloplasty was simulated by reducing annular circumference via displacement-controlled boundary conditions, and Mitralign suture annuloplasty was modeled using discrete nodal constraints. Simulations were performed for prolapse and dilated annulus cases, comparing repairs individually and in combination. Valve competence (regurgitant orifice area, ROA), coaptation/contact area (CA), and leaflet stress and strain distributions were quantified.

\textbf{Results:} In prolapse, TEER restored coaptation but increased leaflet stresses, whereas band and suture annuloplasty produced distinct valve morphologies with lower stress distributions. In dilation, TEER alone left residual regurgitation, while annuloplasty improved closure. Quantitatively, combined TEER + band annuloplasty minimized ROA (0.06 $\text{cm}^2$), maximized CA (2.57 $\text{cm}^2$), and reduced stresses relative to TEER alone, though stresses remained higher than annuloplasty alone.

\textbf{Conclusion:} This study establishes a reproducible, open-source FEM framework for simulating transcatheter TEER and annuloplasty repairs, with the potential to be extended beyond the mitral valve. By quantifying the mechanical trade-offs of TEER, suture annuloplasty, band annuloplasty, and their combinations, this methodology highlights the potential of virtual repair to guide patient selection and optimize surgical planning. }


\maketitle

\noindent 

\section{Introduction}\label{sec:Introduction}
Atrioventricular valve regurgitation (AVVR) is a condition in which the mitral or tricuspid valve fails to close properly during ventricular contraction allowing blood to flow backward from the ventricle into the atrium. AVVR is a major contributor to morbidity and mortality across a broad patient population, affecting both adults with acquired heart disease ~\cite{adult-avvr1, adult-avvr2} and children with congenital heart disease (CHD), particularly those with single-ventricle physiology ~\cite{avv1, avv2, king2019atrioventricular}. Surgical repair remains the standard of care for severe AVVR, but many patients are at elevated risk for open-heart surgery. As a result, transcatheter valve repair techniques have emerged as minimally invasive alternatives. 

A widely adopted technique is transcatheter edge-to-edge repair (TEER) in which a clip is deployed to grasp opposing valve leaflets and improve coaptation, thereby reducing regurgitant flow ~\cite{mitral-teer1,mitral-tri-teer}. TEER has also recently shown similar promising outcomes in pediatric patients with single ventricle heart disease~\cite{jolley2024}. In addition to leaflet-based TEER therapy, transcatheter therapies aimed at annular reduction have been developed for both the mitral~\cite{messika2019} and tricuspid valve~\cite{gray2022}. These innovative therapies emulate the established surgical annuloplasty techniques by~\cite{korber2021} either implanting a band around the annulus to reduce dilation~\cite{cardioband} or by delivering sutures near the commissures to cinch the annulus and enhance leaflet apposition~\cite{mitralign, mitralign2}. These approaches showed successful outcomes, significantly reducing heart failure symptoms and mortality in patients with primary mitral regurgitation~\cite{mack2021}, and lowering hospitalization rates in patients with secondary mitral valve regurgitation related to heart failure~\cite{anker2024, baldus2024, stone2018}. Similar benefits have been observed in tricuspid valve interventions, including reductions in annular dimensions and improvements in functional outcomes such as exercise capacity and quality of life~\cite{sorajja2023, cardioband-tricuspid}.

Patient-specific image-derived modeling, virtual simulations, and physical simulations have informed preprocedural planning, particularly for TEER in CHD patients~\cite{jolley2024, cianciulli2021, ching2023}. However, image-derived modeling does not allow the clinician to preoperatively evaluate the effect of valvular therapies on valve function or the effect of intervention on leaflet mechanics~\cite{narang2021}. Virtual repairs using computational methods emerged as tools to allow surgeons to better understand the effect of interventions on metrics of valve function (e.g. coaptation and regurgitant orifice area) and leaflet mechanics (stress and strain) ~\cite{ge2014virtual, kong2020virtual, choi2014virtual1, choi2017virtual2, rausch-valve-mechanics, wu2022, mechanics-mitral-sun, lee2014}. In addition to the assessment of initial post-repair function, understanding leaflet mechanics may help identify repairs which both improve valve function and are less susceptible to failure over time\cite{sa2023coaptation, zhang2019mechanical}. For example, surgical annuloplasty repairs have been modeled to predict postoperative valve geometry and mechanics, including the application of annuloplasty rings~\cite{choi2014virtual1, kong2018virtual}. More recently, simulations have been extended to application of TEER to both mitral and triscuspid valves~\cite{kong2020virtual, rausch-clip}.

However, a critical clinical question is determining which repair technique (surgical or transcatheter) should be applied to achieve improved and durable valve function. To date no study has compared application of different transcatheter techniques or the combination of techniques as would be desired to determine the optimal repair in an individual patient. For example, there is no current investigation of the effect of clip width and length in clinically available devices on resulting leaflet biomechanics. Similarly, for large coaptation gaps, it is clinically common to employ sequential leaflet capture (grasp one leaflet, move the catheter, then grasp other leaflet). However, the effect of this approach on leaflet function and mechanics has not been previously simulated. Further, TEER simulations to date have made unrealistic assumptions such as fixing the clip in space while clinically the clip moves with the valve leaflets once it is released from the catheter. Finally, most studies to date rely on the use of multiple commercial software platforms, many of which are not fully extensible or modifiable by the end-user to optimize simulations of these complex interventions.

As such, we developed new technical capabilities in our open-source framework~\cite{wu2023, wu2022, maas2012febio, lasso2022slicerheart} to address these challenges. These include a novel implementation of clip release, in which leaflet capture is modeled through sticky contact interfaces and subsequent device–tissue interaction is stabilized by releasing rigid forces and rigid moments. We also incorporated a compliant annulus representation using linear spring elements, allowing physiologic motion of the annular boundary. Importantly, all components of these simulations, from geometry creation to device deployment and valve loading, were performed entirely with open-source software, ensuring transparency, reproducibility, and extensibility. We then applied these capabilities to compare three commercially available transcatheter interventions in a pediatric model of mitral valve disease: ($1$) TEER with a novel “released clip” implementation and evaluation of multiple clip sizes, ($2$) band annuloplasty, and ($3$) annuloplasty using suture-based cinching, as well as combined TEER + annuloplasty repairs. Through these simulations, we demonstrated how different approaches influence leaflet coaptation and valve mechanics, providing a foundation for patient-specific assessment of repair strategies and, in the future, improved identification of the optimal intervention for individual patients.

\section{Methods}\label{sec:Methods}

\begin{figure}[b]
    \centering
    \includegraphics[width=\textwidth]{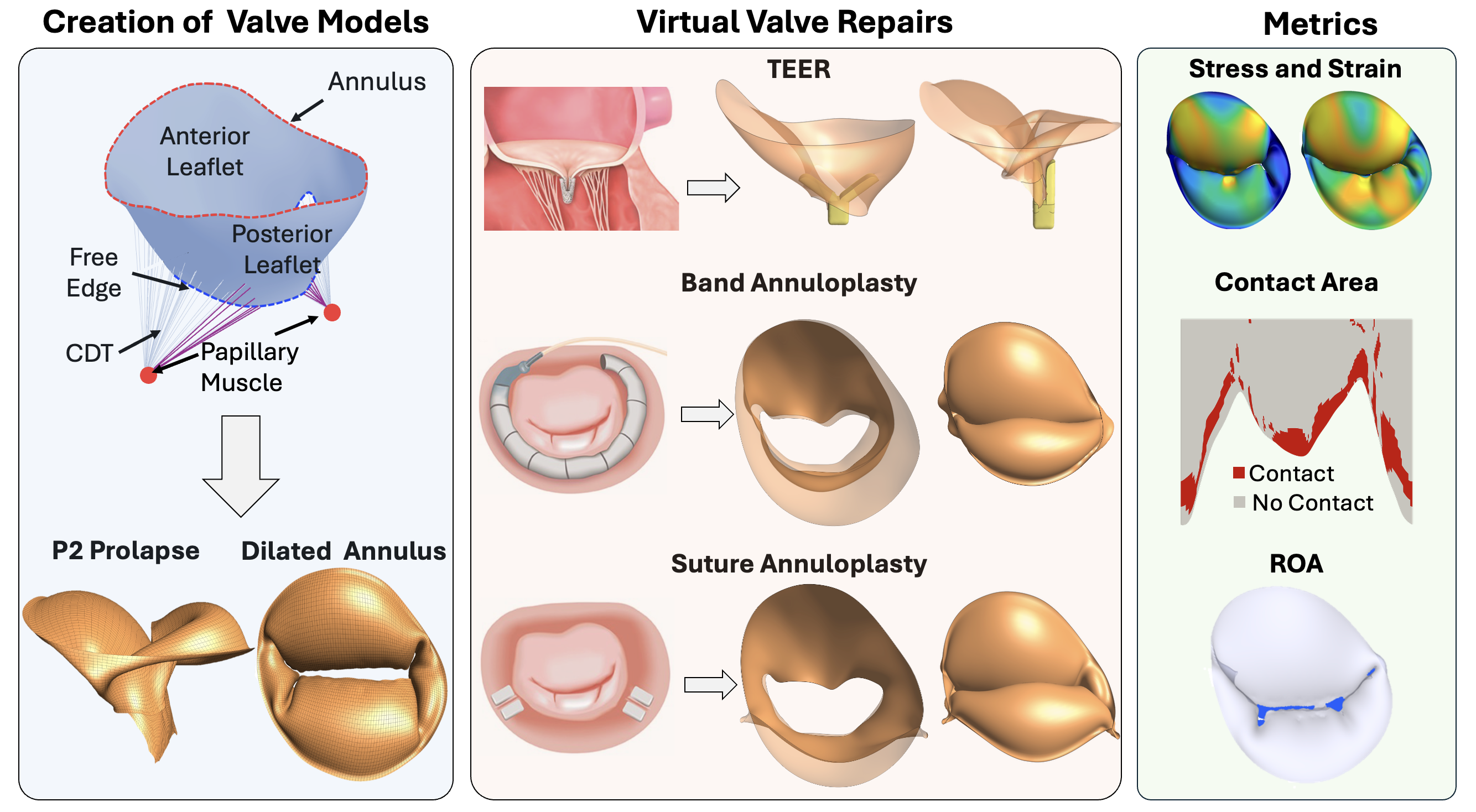}
    \caption{ Valve model generation, virtual valve repairs using FEM, and outcome metrics. Left: A 3D mitral valve was reconstructed from mid-diastolic echocardiographic images (workflow described in~\cite{wu2022}). From this baseline model, two regurgitant phenotypes were created: P2 prolapse and annular dilation. Middle: Clinical repair illustrations are paired with the corresponding FE models that mimic each procedure: TEER (e.g., MitraClip, adapted from Abbott’s original site: www.mitraclip.com), band annuloplasty (e.g., Cardioband by Edwards Lifesciences, adapted from \cite{gasior2019direct}), and suture annuloplasty (e.g., Mitralign by Mitralign Inc., adapted from \cite{gasior2019direct}). Right: Post-repair outcomes quantified from the FEM include leaflet stress/strain distributions, leaflet coaptation/contact area (CA), and regurgitant orifice area (ROA) shown in blue.}
    \label{fig:workflow}
\end{figure}

We integrated new features into our established open-source pipeline~\cite{wu2023} using SlicerHeart~\cite{lasso2022slicerheart}, 3D Slicer~\cite{fedorov20123d}, and FEBio~\cite{maas2012febio} to perform virtual transcatheter valve repairs in a image-derived model of a pediatric mitral valve. The study workflow is summarized in Figure~\ref{fig:workflow}.

A patient-specific mitral valve model was reconstructed from $3$D echocardiographic images (Figure~\ref{fig:workflow}, left panel) as described in~\cite{wu2023}, and corresponding regurgitant models were derived from this baseline geometry. All FEBio models are available in the FEBio Model Repository, accessible through FEBio Studio. Finite element (FE) simulations were then conducted for three virtual repair strategies: TEER with clip deployment, band annuloplasty, and suture-based annuloplasty (Figure~\ref{fig:workflow}, middle panel). From the simulation results, we extracted key biomechanical metrics including stress and strain distributions, leaflet contact area (CA), and regurgitant orifice area (ROA) (Figure~\ref{fig:workflow}, right panel) as previously described~\cite{wu2023}. Each step of this workflow is described in detail in the following sections.

\subsection{Creation of Valve Geometry}

An image-informed mitral valve model derived from a $15$-year-old male with normal cardiac anatomy and no mitral regurgitation~\cite{wu2022} was used to generate a posterior leaflet prolapse and annular dilation models. This dataset was selected because it represents a physiologically normal mitral valve, providing a representative baseline anatomy for simulating regurgitant conditions. The use of patient data was approved by the Institutional Review Board at the Children’s Hospital of Philadelphia.

The $3$D mitral valve geometry (Figure~\ref{fig:workflow}, left) included anterior and posterior leaflets, annulus edge, free edge, papillary muscles, and the chordae tendineae (CDT). The CDT were created and attached to the valve surface following the procedure detailed in~\cite{wu2023} based on the work by Khalighi \textit{et al}.~\cite{khalighi2017mitral}. 

The posterior leaflet prolapse was created by decreasing the tension of the posterior CDT (purple chords in~\ref{fig:workflow}, left) to allow the posterior leaflet to billow beyond the anterior leaflet, creating a regurgitant gap exceeding $0.4$,cm$^{2}$, the threshold for severe mitral regurgitation~\cite{zoghbi2020recommendations}. Similarly, to create an annular dilation case, the diameter of the annulus curve was increased by $2.8\%$ without lengthening the leaflets to allow a regurgitant gap. These modifications were performed using open-source platforms SlicerHeart, $3$D Slicer, and FEBio as detailed in~\cite{wu2023}. Two regurgitant valve models, P2 prolapse and dilated annulus, were included in FEBio repository (). 

The final regurgitant P2 prolapse and dilated annulus models were illustrated in Figure~\ref{fig:workflow}, left panel.

\subsection{Virtual Valve Repairs}
\label{FEM}

The virtual valve repairs were simulated to evaluate the effects of different transcatheter interventions. Each clinical repair strategy was paired with a corresponding FE model designed to mimic the procedure, as illustrated in Figure~\ref{fig:workflow} (middle panel). The top row depicts a transcatheter edge-to-edge repair (TEER, e.g., MitraClip), in which the clip arms grasp the mitral leaflets and the post-repair configuration was assessed under systolic pressurization. The middle row illustrates a band annuloplasty procedure (e.g., Cardioband), modeled by displacing selected annular nodes to replicate the geometric remodeling produced by the implant; the resulting FEM-predicted valve shape under pressure is shown. Finally, the bottom row shows a suture annuloplasty procedure (e.g., Mitralign), simulated by virtual sutures at the annulus.

The mitral valve CDT were modeled as tension-only, two-node linear springs connecting leaflet insertion points to papillary muscle tips as previously described~\cite{wu2023}. A force–displacement behavior was implemented in FEBio \cite{maas2023febio-material}, with zero tension below a displacement threshold and linear response beyond it. The effective nonlinear spring force was defined as $F(x)=f_{s}$\texttimes$G(x)$, where $f_{s}$ is the user-defined spring force of $60$ \text{mN}, $G(x)$ is a unitless scaling function that defines the force–displacement behavior, and $x$ is the change in spring length. The force–displacement scaling function was formulated as

\begin{equation}
G(x) = \begin{cases}
0, & x \leq \delta \\
x - \delta, & x > \delta
\end{cases}
\end{equation}

\noindent Here, $\delta$ represents user-defined displacement-threshold which was set to $5$\,mm~\cite{wu2023}. Chordal parameters were determined through an iterative procedure, gradually increasing the spring tension until equilibrium was reached, as described in \cite{wu2022,mangine2024effect}. In order to create the prolapsed valve, $f_{s}$ was decreased to $30$\,mN for the CDT (depicted as purple lines in Figure~\ref{fig:workflow}, bottom left) in the P$2$ region of the leaflet~\cite{wu2022}.

We adopted the incompressible, isotropic, hyperelastic Lee-Sacks~\cite{lee2014} constitutive model to represent the mitral valve leaflet tissue, with the strain energy density function defined as

\begin{equation}
\label{Eq:LeeSacks}
    \psi = \frac{c_0}{2}\left(I_{1}-3\right) + \frac{c_1}{2}\left(e^{c_{2}\left(I_{1}-3\right)^2}-1\right)
\end{equation}

\noindent Where, $c_0$ ($200$\,MPa), $c_1$ ($2968.4$\,MPa), and $c_2$ ($0.2661$) are material coefficients, $I_1$ is the first invariant of the right
Cauchy-Green deformation tensor $\mathbf{C}=\mathbf{F}^T\mathbf{F}$. Near-incompressibility was enforced using a bulk modulus of $5000$\,kPa.

The valve surface was discretized using $4$-node linear quadrilateral (Quad4) shell elements with an assumed thickness of $0.396$\,mm~\cite{wu2023, mangine2024effect}. For the initial simulations, pinned boundary conditions were applied at the annular edge and papillary muscle tips. A systolic pressure of $100$\,mmHg was applied to the leaflet surfaces to induce valve closure. Leaflet contact was modeled using a potential-based contact formulation~\cite{kamensky2018contact,wu2022,wu2023,laurence2024}. For the leaflets, the contact potential was defined with a penalty stiffness multiplier ($k_{c}=2$) and a steep force exponent ($p=4$), producing a strong rise in repulsion as leaflets approached. Interaction forces were active when leaflets were within $0.5$\,mm ($R_{\text{out}}$) and increased fully below $0.2$\,mm ($R_{\text{in}}$). To improve robustness, a facet-based filtering parameter ($w_{\text{tol}} = 0.6$) was applied to prevent neighboring facets from being considered contact pairs. This allowed the leaflets to slide and coapt realistically during opening and closing without artificial sticking, while maintaining stability prior to clip capture. Implicit dynamic analysis was carried out using the generalized-alpha integration scheme with mass damping \cite{febio-bodyloads} applied to the leaflets.

Following this FEM methodology, we simulated three virtual repair strategies: TEER, band annuloplasty, and suture-based annuloplasty. The specific implementation of each repair is described in the following subsections. 

\subsubsection{Transcatheter Edge-to-Edge Repair} \label{Transcatheter Edge-to-Edge Repair}

\textbf{MitraClip Modeling and Deployment Protocol}

\begin{figure}
    \centering
    \includegraphics[width=\textwidth]{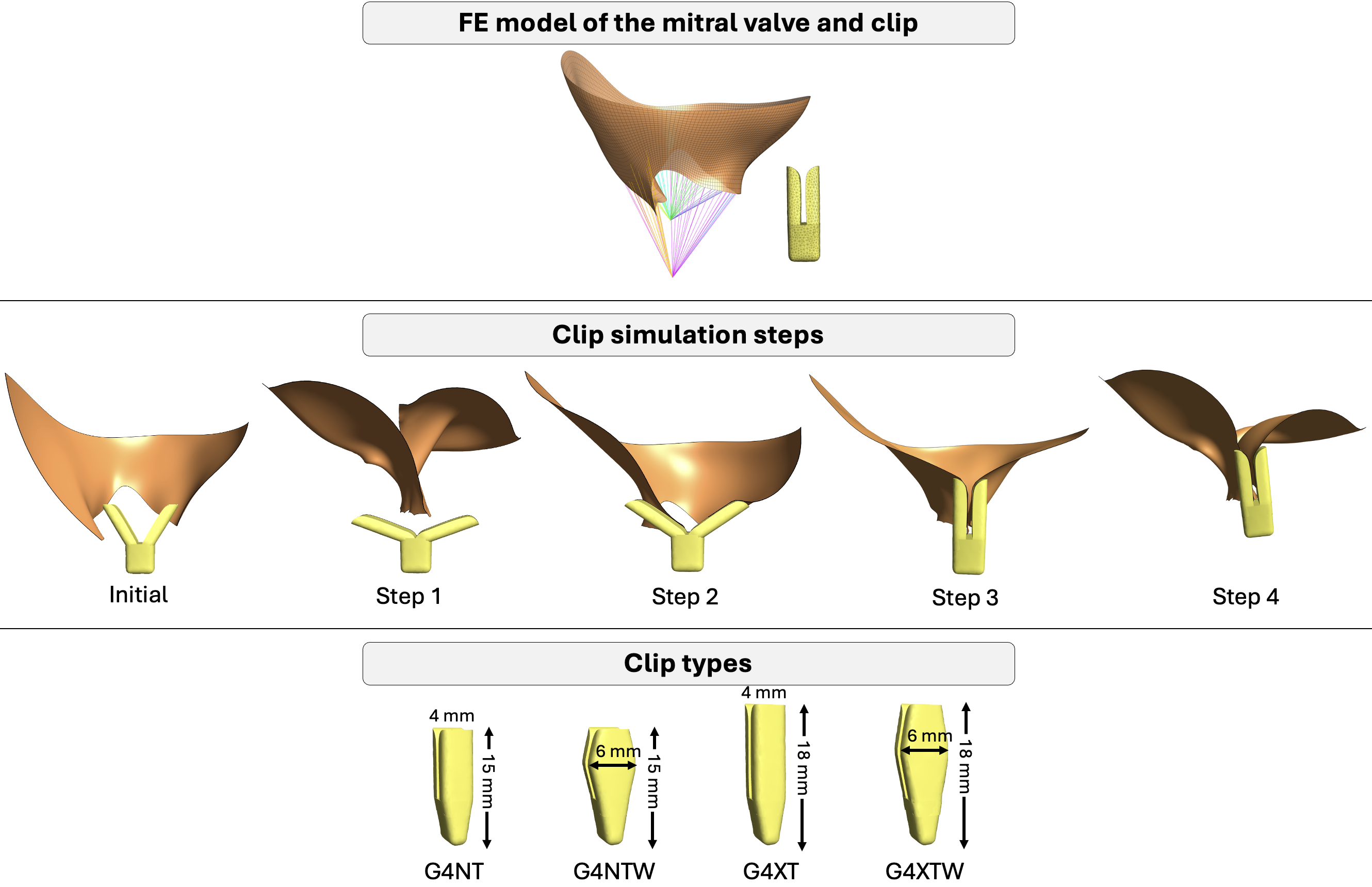}
    \caption{TEER clip insertion and clip types. (a) FE model of the mitral valve with chordae tendineae and the transcatheter clip. (b) Stepwise simulation of clip deployment, showing leaflet capture and coaptation over sequential steps. (c) Commercial clip types used in virtual TEER simulations, including G4NT, G4NTW, G4XT, and G4XTW, each defined by distinct arm lengths and widths.}
    \label{fig:clip-steps}
\end{figure}

The detailed steps of MitraClip procedure were shown in Figure \ref{fig:clip-steps}. Clip models for the base and arms of four MitraClip G$4$ clips (Abbott Cardiovascular, Plymouth, MN; sizes: NT, NTW, XT, XTW) were adapted from the ValveClip Device Simulator module~\cite{cianciulli2021} in the SlicerHeart extension for 3D Slicer (www.slicer.org). Each clip was meshed with 4-node tetrahedral elements in FEBio Studio. The clip was represented as three rigid bodies: two arms (``Clip-Anterior'' and ``Clip-Posterior'') and a base (``Clip-Base''). The FE models of the valve and clip were illustrated in Figure \ref{fig:clip-steps}, top.

\textbf{Initial Step:} Before simulating pressure loading, the clip was positioned at the site of prolapse based on the closed valve reference geometry. The anterior and posterior arms, along with the base, were assigned rigid body definitions, and their degrees of freedom were constrained to maintain alignment with the mitral valve. Translational (Rx, Ry, Rz) and rotational (Rv, Rw) degrees of freedom were fixed, leaving only Ru free to allow controlled arm rotation. This initialization ensured that the clip began in an anatomically appropriate position relative to the valve and that subsequent opening and closure were driven in a physiologically consistent manner.

This initial configuration and subsequent steps of TEER are summarized in Figure \ref{fig:clip-steps}, middle row.

\textbf{Step $1$: Valve pressurization with clip arms opening.}
The valve leaflets were loaded with physiological pressure ($100$\,mmHg) using a prescribed load curve, which drove the valve into its closed systolic configuration. Simultaneously, the clip arms were rotated outward about their local $R_{u}$ degree of freedom by $\pm 1.22$ rad ($70^{\circ}$), controlled by a prescribed load curve. All other translational ($R_{x}, R_{y}, R_{z}$) and rotational ($R_{v}, R_{w}$) degrees of freedom were fixed, leaving only $R_{u}$ free for the arm rotation. At this stage, leaflet deformation was governed solely by pressure, while the clip arms transitioned to the fully open configuration.

\textbf{Step $2$: Valve opening and leaflet entry between the clip arms.}
As the pressure load curve drove the valve into its diastolic opening phase, the anterior and posterior leaflets began to move between the open clip arms.

\textbf{Step $3$: Clip closure and leaflet grasp.}
With the leaflets positioned between the open arms, the clip arms were rotated inward about their $R_{u}$ degree of freedom following the prescribed load curve ($\pm 1.22$ rad, $70^{\circ}$). At this stage \textit{in vivo}, a gripper drops pinning the leaflets between the clip and gripper. The simultaneous (dual) leaflet capture was enforced without the gripper, using \texttt{sticky contact} interfaces defined for the anterior and posterior clip arms. Sticky contact was introduced with the adhesion penalty ($\epsilon = 1000$) scaled by a prescribed load curve. The load curve was defined to ramp up almost immediately after the start of Step 3 (onset at $t = 0.001$), ensuring stable activation of adhesion as soon as the leaflets contacted the clip arms.

The sticky contact used a penalty-based enforcement (\texttt{laugon = PENALTY}) with a penalty factor of $1000$, balanced to prevent excessive penetration while maintaining convergence. A very tight contact tolerance ($1\times 10^{-6}$) and search tolerance ($1\times 10^{-4}$) ensured accurate detection of leaflet--clip engagement, while the gap offset ($0.2$\,mm) set the initial activation distance for adhesion. The option \texttt{flip\_secondary = 1} is necessary to make sure the surfaces adhere to the orientation convention that is assumed in contact in FEBio.

This setup ensured that as soon as leaflet--clip contact occurred during closure, the leaflets adhered irreversibly to the clip surfaces, replicating the mechanical grasp of TEER deployment.

\textbf{Step $4$: Release of deployment controls and physiologic pressurization.}
In the final step, the prescribed rigid rotations and fixation constraints that were used to control clip opening, positioning, and closure during deployment were released. This was implemented in FEBio by driving all rigid forces and rigid moments ($R_{x}, R_{y}, R_{z}, R_{u}, R_{v}, R_{w}$) acting on the clip base and arms to zero using \texttt{rigid\_force} and \texttt{rigid\_moment} definitions with \texttt{TARGET} load type \cite{febio-rigidloads-47}. In FEBio, the \texttt{TARGET} specification ramps loads linearly from their initial values to the target value, in this case zero, so that the clip transitioned smoothly from externally controlled motion to a fully leaflet tissue-coupled state without instability.

At the same time, the anterior and posterior arms were rigidly coupled to the base using \texttt{rigid\_connector} constraints of type \textit{rigid lock} \cite{febio_materials_34}. These connectors eliminated relative motion between the arms and base, enforcing closure of the device as a single rigid body. Stability was ensured using a gap tolerance ($1\times 10^{-4}$), angular tolerance ($1\times 10^{-4}$), and force/moment penalties of $100$. The option \texttt{auto\_penalty = 1} was also enabled, which scales the penalty factor automatically based on element size and material properties, while the user-defined penalty value serves as a unitless scaling factor.

After these adjustments, physiological systolic pressure was reapplied to the atrial leaflet surfaces via a prescribed load curve. Because the clip was internally locked but no longer fixed to the global reference frame, the entire clip--leaflet complex was free to translate, rotate, and deform together with the valve. This represents a key novelty of the FEBio implementation: following leaflet capture, the device was not artificially held in space, but instead moved dynamically with the valve under physiologic loading, stabilized entirely by leaflet contact, tissue tension, and pressure balance.

\textbf{Pinned and Compliant Annulus}: Next, we compared two annular boundary conditions: a rigid (pinned) annulus and a compliant annulus modeled using linear springs, allowing for dynamic expansion and contraction during pressurization. Our compliant annulus model (implemented in FEBio), is based on the dynamic annulus approach described by Haese \textit{et al}.~\cite{Haese2025Tricuspid}, where linear springs are connected radially to the annulus and non-annular nodes are pinned, enabling deformation without fixed boundary constraints. In their study on the tricuspid valve, the authors demonstrated that varying the spring stiffness significantly impacts stress and strain distributions. Their results were highly sensitive across a broad stiffness range ( \(0.01\text{--}1\,\mathrm{N/mm}\) ) but for \(k \gtrsim 0.5\,\mathrm{N/mm}\), the responses became similar to those of a pinned annulus. Because the mitral annulus is stiffer than the tricuspid annulus~\cite{yucel2020tricuspid}, we chose an intermediate value of \(k = 0.3\,\mathrm{N/mm}\): high enough to reflect the greater mitral stiffness yet low enough to remain in the sensitivity regime identified by Haese \textit{et al}.~\cite{Haese2025Tricuspid}.

\textbf{Clip Device Length and Width:} Commercial G4 clip devices are available in multiple lengths and widths (Figure~\ref{fig:clip-steps}, bottom). Selecting the appropriate device requires consideration of patient-specific anatomy and valve size. In some cases, more than one clip may appear to provide an adequate anatomical fit. However, understanding how different clip sizes influence mechanical metrics is essential, and FEM offers a means to evaluate these effects during procedure planning. Therefore, multiple sizes of MitraClip G4  (Figure~\ref{fig:clip-steps}, bottom) were modeled and simulated in the prolapse case, and the resulting metrics were systematically compared.

\textbf{Staged Leaflet Capture:} We simulated a sequential leaflet capture technique, which may be beneficial in cases where the leaflets are too far apart to be grasped simultaneously. In such scenarios, clinicians may opt to capture one leaflet first, followed by the second. In staged leaflet capture simulation, during analysis step $2$ (described in Figure \ref{fig:clip-steps}, middle row, the sticky contact constraint was first applied to the anterior arm of the clip, instead of both arms. A new analysis step was added that pressurized the valve and prescribed a new location to the clip. Next, sticky contact was re-activated for the posterior arm to capture the posterior leaflet at mid-diastole. Lastly, analysis was continued following Step $3$ and $4$. We hypothesized that this process enables the clip to capture a larger portion of the posterior leaflet compared to capturing it simultaneously. The video of staged leaflet capture was included in Online Resource $1$.

\subsubsection{Band Annuloplasty} \label{Annuloplasty}
\begin{figure}
    \centering
    \includegraphics[width=\textwidth]{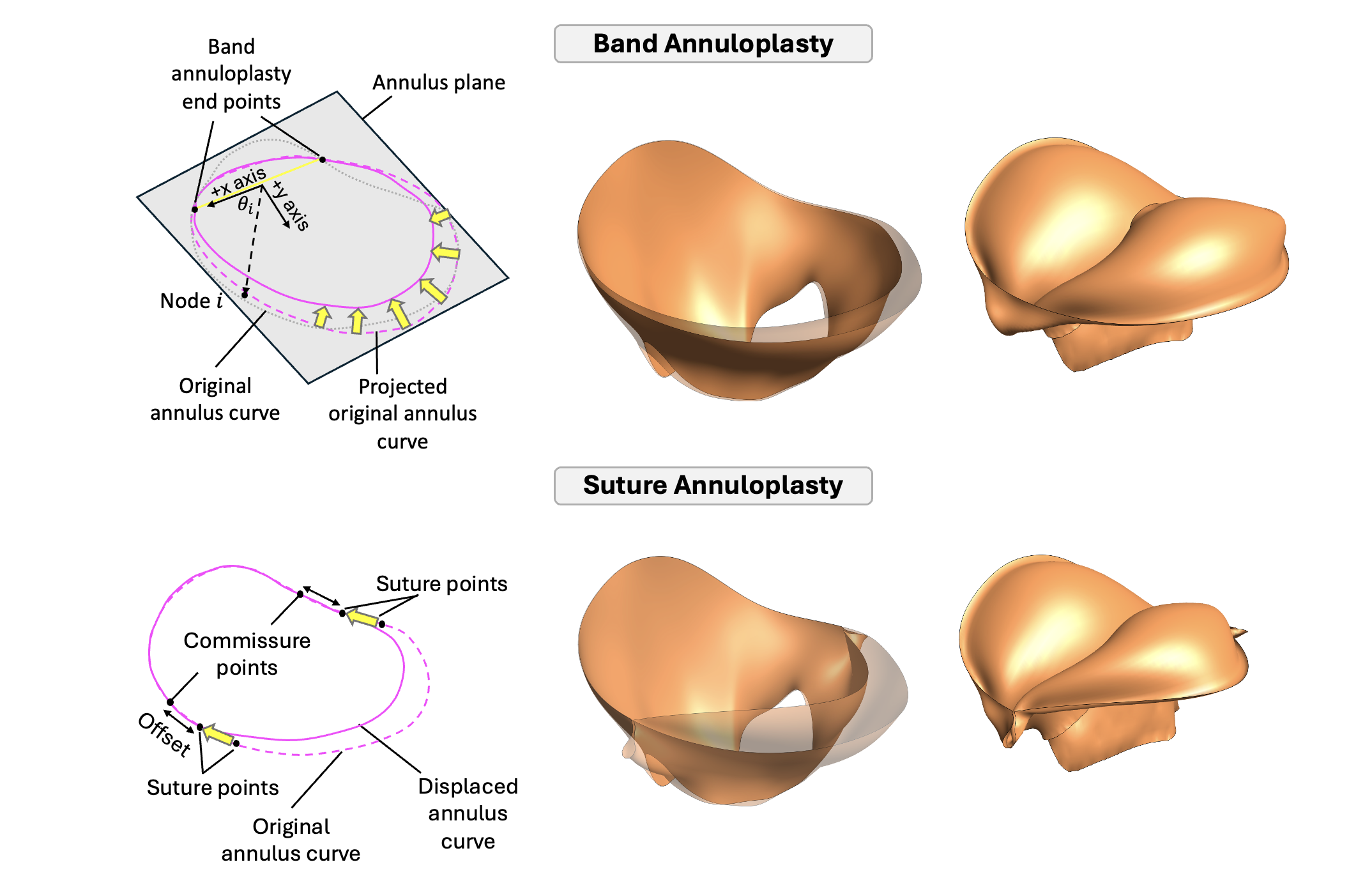}
    \caption{Virtual modeling of band annuloplasty and commissuroplasty procedures. Top row: Schematic of band annuloplasty, where annular nodes between two endpoints are displaced inward to reduce annular circumference, along with resulting FE simulation of the repaired valve. Bottom row: Schematic of suture-based commissuroplasty, in which selected commissural regions are cinched via paired suture points to modify annular shape, followed by the resulting valve deformation. Both techniques were implemented through displacement-based boundary conditions applied to annular nodes.}
    \label{fig:annuloplasty}
\end{figure}

A ring based transcatheter annuloplasty, mimicking the CardioBand (Edwards Lifesciences, Irvine, CA, USA) technique, was modeled by decreasing the annulus circumference by applying displacement to selected annulus nodes using a custom-developed Python script. The annulus nodes were projected onto a $2$D plane and the endpoints of the annuloplasty band, together with the intervening annulus nodes, were fixed (Figure~\ref{fig:annuloplasty}, top left). The remaining annular nodes, indicated by the dashed curve, were displaced inward according to the displacement function:

\begin{equation}
\label{Eq:Band Annuloplasty Node Displacements}
    \begin{Bmatrix}
\mathrm{u}_{i} \\
\mathrm{v}_{i}\\
\mathrm{w}_{i}
\end{Bmatrix}
=
\begin{Bmatrix}
a\mathrm{x}_{i} sin{\theta}_{i} \\
a\mathrm{y}_{i} sin{\theta}_{i} \\
0
\end{Bmatrix}
\end{equation}

\noindent Here, $a<1.0$ is the percent decrease in annular circumference, ${x}_{i}$ and ${y}_{i}$ are the $x$ and $y$ coordinates of the annulus node $i$. ${\theta}_{i}$ is the angle between the $x$-axis and the origin to node $i$ vector. This was accomplished in analysis step $1$, and once the nodes were displaced along the annulus plane, the valve leaflets were pressurized as described in Section~\ref{FEM} and analysis were completer. The virtual band annuloplasty procedure is illustrated step-by-step in Figure~\ref{fig:annuloplasty}.

\subsubsection{Suture Annuloplasty} \label{Commissuroplasty}

To simulate the suture procedure that mimics the Mitralign system (Mitralign Inc., Tewksbury, MA, USA.), we created a custom Python script that modified the mitral valve annulus geometry based on user-defined inputs. Commissure points, which served as references for suture placements, were first identified in Slicer3D. An offset value, defined as the number of nodes away from the commissure, was specified to locate the starting suture point. The ending suture point was then determined based on a user-defined suture length. A displacement was applied to the ending suture node, directing it toward the starting suture node and stopping $0.4$\,mm short to account for leaflet thickness. Nodes between the two ending suture points were passively displaced as a result of the annular cinching. To preserve local annular spacing, a linear constraint boundary condition was applied, preventing changes in distance between adjacent annular nodes. The suture annuloplasty was performed following the settings as described in Section~\ref{FEM}. The virtual suture annuloplasty procedure was illustrated in Figure \ref{fig:annuloplasty}, bottom row.

\subsection{Analysis of Mechanical and Functional Metrics}

We evaluated the effects of transcatheter mitral valve repairs using both functional and mechanical metrics. For functional assessment, we focused on clinically relevant parameters such as ROA and leaflet CA, as these are key indicators of valve competence after repair~\cite{rivera1994effective, benfari2018mitral}.
We used a fully-automated method to accurately quantify the ROA by coupling a shrink-wrapping method with raycasting~\cite{wu2023}. The CA was calculated by the surface area of the elements with non-zero traction applied by the contact algorithm~\cite{kamensky2018contact}.

To evaluate mechanical performance, we calculated and compared first principal stresses and strains in the mitral valve leaflets based on its relationship with valve remodeling~\cite{howsmon2020mitral}. These metrics are critical because elevated stress and strain concentrations have been associated with an increased risk of rupture~\cite{rausch2011characterization}, which is a potential cause of mitral valve repair failure. Prior studies have emphasized that achieving an optimal distribution of mechanical stresses may enhance valve durability and long-term function by reducing the likelihood of tissue fatigue and mechanical failure~\cite{rausch2011characterization, prot2010modelling}.

\section{Results}\label{sec:Results}

Using FE simulations, we quantified leaflet stress and strain, CA, and ROA before and after each repair, and compared repair strategies across these metrics. To ensure robust comparisons and reduce the impact of localized numerical artifacts, leaflet stresses and strains were reported as 95th percentile values rather than absolute maxima in the following subsections.

\subsection{Effect of TEER}

Below, we first assessed how different simulation scenarios and boundary conditions influence TEER. Then presented the primary analysis of TEER effects for prolapse case.

\begin{figure}[h!]
    \centering
    \includegraphics[width=0.75\textwidth]{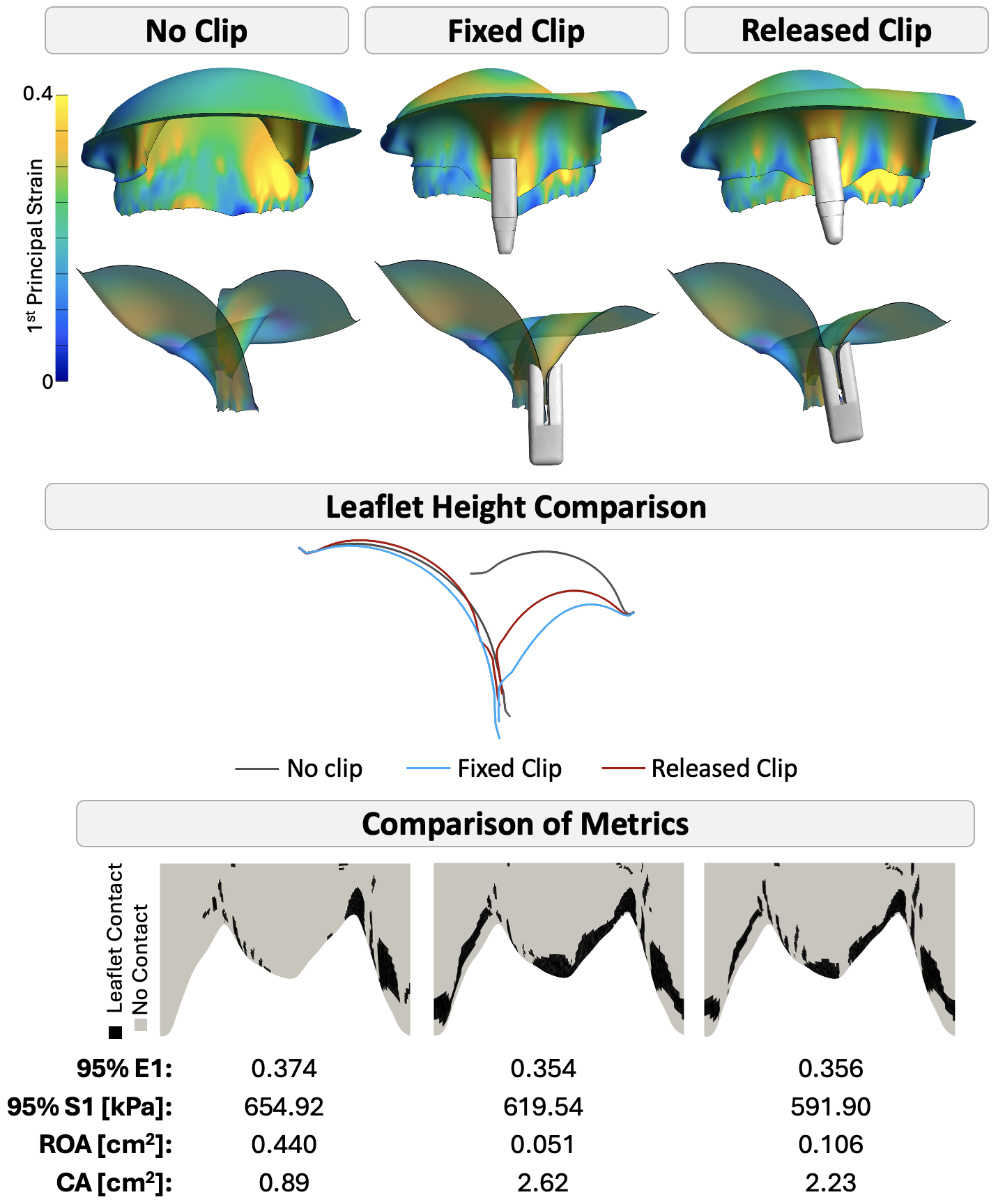}
    \caption{FE simulations of mitral valve repair with no clip, fixed clip, released, and staged clip configurations. Top: first principal strain distributions across valve leaflets. Middle: valve leaflet morphology illustrating coaptation height differences, with leaflet height comparison shown as overlaid profiles (black: no clip, orange: fixed clip, blue: released clip). Bottom: leaflet contact maps showing contact regions (black) versus no contact (gray). Quantitative metrics are summarized below, including 95\%ile of first principle strain (E1) and stress (S1), ROA, and CA.}
    \label{fig:fixed-vs-released}
\end{figure}

\textbf{Clip Release vs. Fixed Clip}: The outcomes of the released clip simulations were compared against both the regurgitant valve (prolapse case) and a simplified “fixed clip” configuration used as a reference. Figure~\ref{fig:fixed-vs-released} illustrates the distribution of first principal strain, leaflet height profiles, and quantitative metrics. Without a clip, the valve exhibited prolapse with poor coaptation and elevated strain. The fixed clip restored coaptation and substantially reduced strain and stress, but the device remained artificially constrained. By contrast, the released clip preserved coaptation while permitting the device to move with the leaflets during pressurization, which altered strain distributions and peak stresses relative to the fixed case.

Leaflet height analysis showed that both fixed and released clips reduced billowing compared to the prolapsed valve, with the released configuration more closely approximating physiologic closure. Quantitatively, the released clip achieved a slightly larger CA and significantly lower stresses than the fixed clip, while maintaining comparable ROA and strain levels (95\% E1 = 0.356 vs. 0.354). The discrepancies between the results underscore that incorporating clip release yields more physiologically realistic valve–device interactions, justifying its use in subsequent analyses.

\textbf{Pinned and Compliant Annulus}: The comparison of metrics obtained from pinned and compliant annulus conditions is presented in Table.~\ref{tab:metrics}. Specifically, we compared the pinned annulus case with a compliant annulus model using a stiffness value of $k=0.3\,\mathrm{N/mm}$ for the linear springs at the annulus nodes. To assess sensitivity, we further varied the spring stiffness between $0.15$ and $0.5\,\mathrm{N/mm}$ and compared these results against the pinned case (Appendix B, Figure ~\ref{fig:compliant annulus appendix}). Across all compliance assumptions tested for this mitral valve model, differences in the evaluated metrics were negligible. Accordingly, a pinned annulus model was adopted for the remainder of this study.

\begin{table}[h!]
\centering
\begin{tabular}{lcc}
\hline
 & \textbf{Pinned Annulus} & \textbf{Compliant Annulus} \\
\hline
\textbf{95\%ile E1} & 0.356 & 0.361 \\
\textbf{95\%ile S1 [kPa]} & 591.90 & 640.50 \\
\textbf{ROA [cm$^{2}$]} & 0.10 & 0.12 \\
\textbf{CA [cm$^{2}$]} & 2.23 & 2.00 \\
\hline
\end{tabular}
\caption{Comparison of metrics between the pinned and compliant annulus using linear spring constant k=0.3 N/mm.}
\label{tab:metrics}
\end{table}

\textbf{Effect of Device Length and Width}

Based on our investigation of different boundary conditions and valve repair scenarios, we performed TEER simulations under a pinned annulus condition and allowing the clip to move with the valve leaflets during pressurization. 

\begin{figure}[h!]
    \centering
    \includegraphics[width=\textwidth]{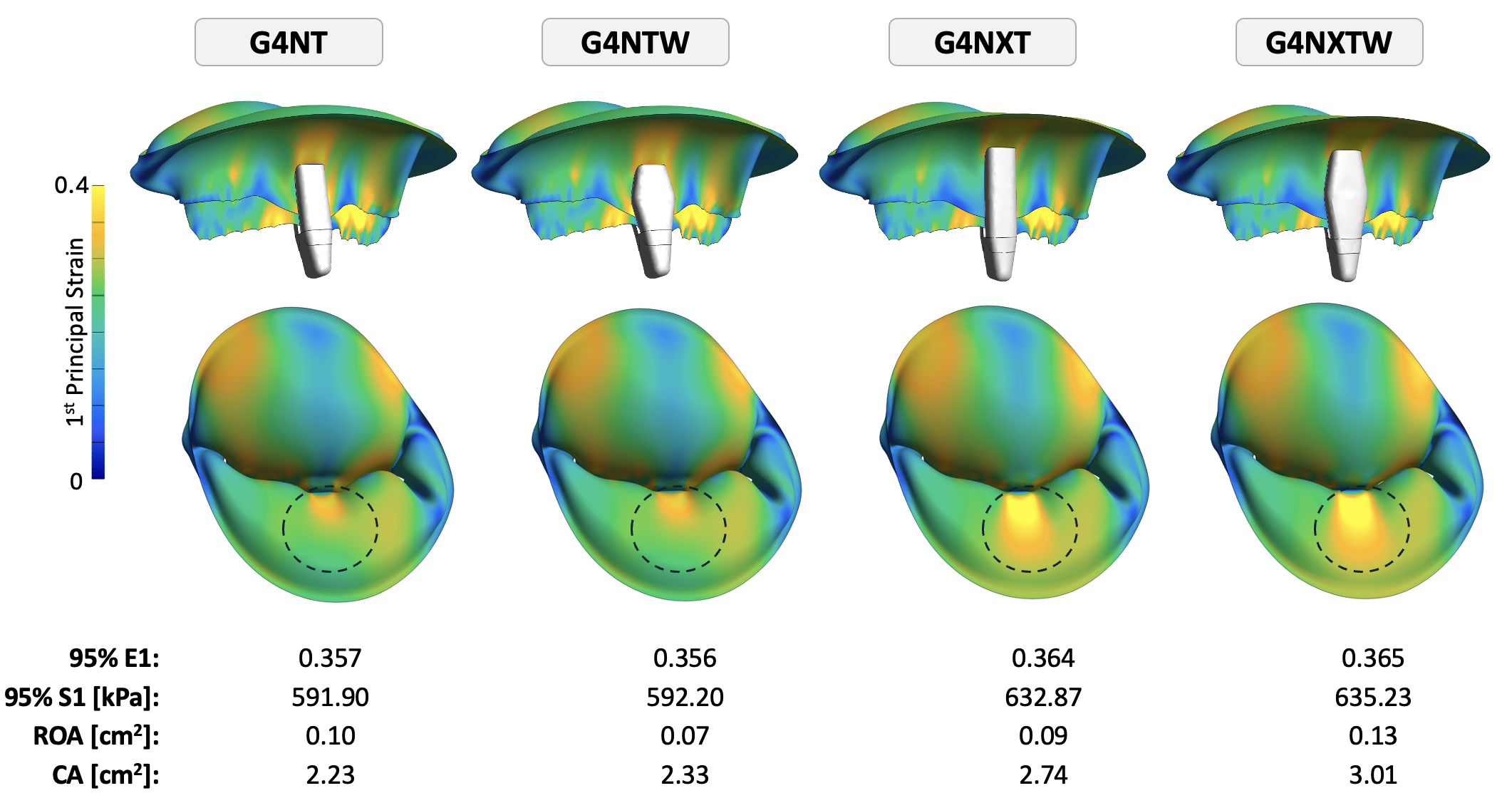}
    \caption{Comparison of mitral valve mechanics across different clip geometries. Strain distributions and coaptation maps illustrate device-specific differences, while quantitative metrics (strain, stress, ROA, CA) highlight variations in repair performance.}
    \label{fig:clip type}
\end{figure}

We simulated TEER with four clip geometries: G4NT, G4NTW, G4XT, and G4XTW. The results are summarized in Figure~\ref{fig:clip type}. Across the four clip configurations, all devices restored leaflet coaptation and substantially reduced ROA compared to the prolapsed baseline valve. However, differences in leaflet mechanics and stress–strain distributions were evident between clip types. The G4NT and G4NTW configurations demonstrated relatively uniform strain patterns with lower stress magnitudes, suggesting more balanced load distribution. In contrast, larger clips, G4XT and G4XTW, produced localized increases in both strain and stress (S$1$: $592\,kPa$ for NT/NTW vs. $632$–$635\,kPa$ for XT/XTW) on the leaflets, near the clip arms (circled in dashed black), reflecting the stiffer geometry and broader grasp of these devices. While these configurations provided larger coaptation areas, they also imposed higher leaflet stresses, which may influence long-term durability. Although there were some differences in the measured metrics, each clip appeared to achieve successful valve closure under pressurization for this prolapse case. 

\begin{figure}[h!]
    \centering
    \includegraphics[width=\textwidth]{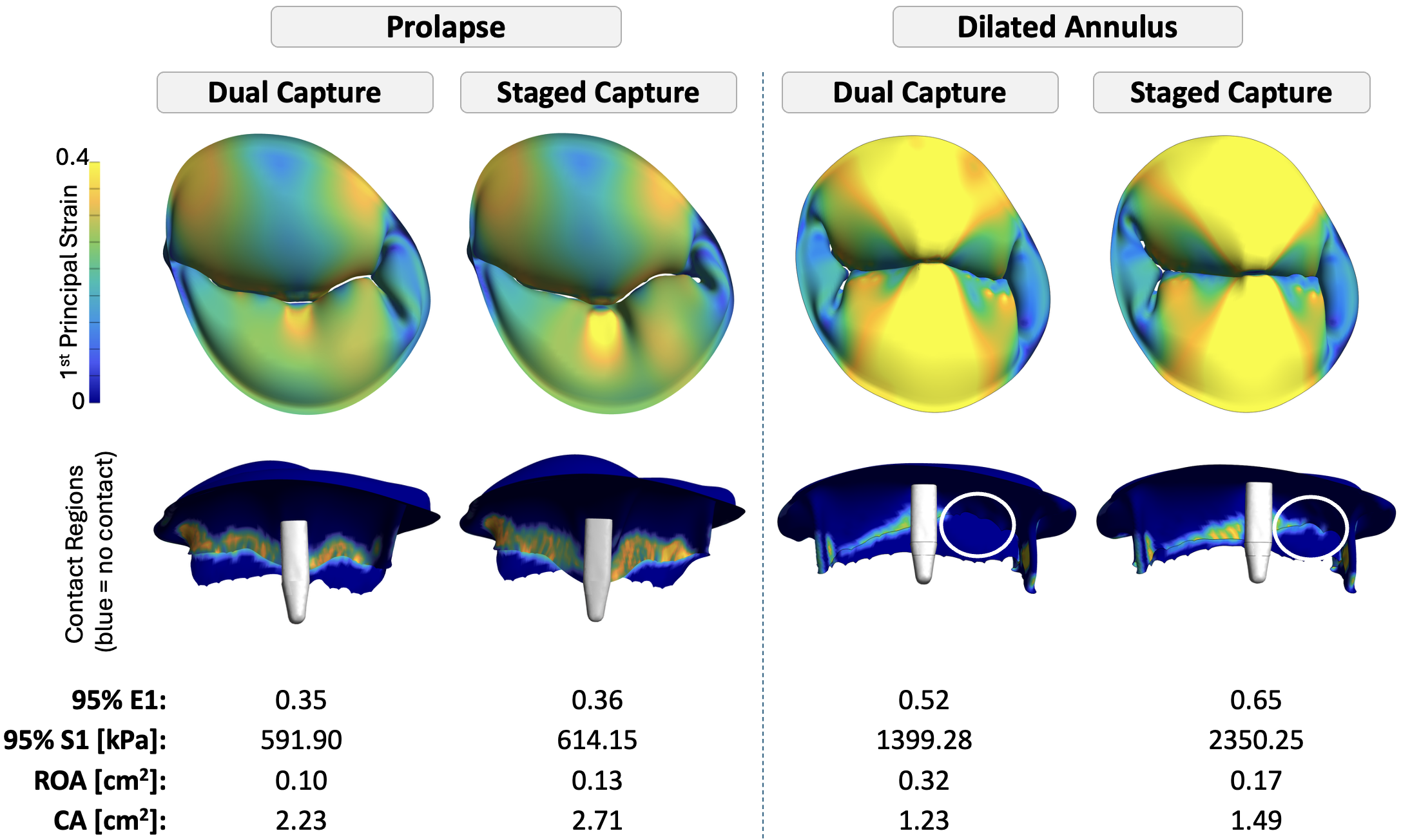}
    \caption{Comparison of dual versus staged leaflet capture in TEER across prolapse and dilated annulus models. Top: distributions of first principal strain on the mitral leaflets. Middle: leaflet–clip contact regions (blue = no contact). Bottom: Quantitative metrics are summarized below, including 95\%ile of first principle strain (E1) and stress (S1), ROA, and CA.}
    \label{fig:staged capture}
\end{figure}

\textbf{Effect of Staged Sequential Leaflet Capture}: TEER with the G4NT clip was simulated in both prolapse and dilated annulus models, comparing simultaneous (dual) and staged leaflet capture strategies (Figure~\ref{fig:staged capture}). In the prolapse model, both dual and staged capture achieved complete leaflet coaptation, eliminating regurgitation. Staged capture increased the CA (2.71 cm² vs. 2.23 cm²) compared with dual capture, reflecting improved engagement of the posterior leaflet. However, this benefit came at the expense of elevated leaflet stresses localized at the grasping sites (95\% S1: 614 kPa vs. 592 kPa), relative to dual capture.

In the dilated annulus model, dual capture successfully secured the anterior leaflet, but the enlarged annular dimension and shorter posterior leaflet prevented complete capture during simultaneous closure, leaving one side partially regurgitant ((ROA: 0.32 cm²) (circled in Figure~\ref{fig:staged capture}, right). By contrast, staged capture enabled better engagement of the posterior leaflet, which improved overall leaflet CA (1.49 cm² vs. 1.23 cm²) and reduced ROA (0.17 cm² vs. 0.32 cm²). These benefits, however, were offset by substantially increased leaflet stresses (95\% S1: 2350 kPa vs. 1399 kPa) and strain (95\% E1: 0.65 vs. 0.52), indicating a significant mechanical burden when staged capture was used in the context of annular dilation. These findings highlight the importance of considering alternative or complementary repair strategies, such as annuloplasty, which can be evaluated alongside TEER in both prolapse and dilation scenarios.

\subsection{Effect of Band Annuloplasty}

We first investigated band annuloplasty in the prolapse model by simulating annular reductions from 5\% to 30\% (Figure ~\ref{fig:Annuloplasty Repairs}). With increasing reduction, leaflet coaptation improved progressively, and complete closure was achieved at approximately 30\% reduction. Below this threshold, residual ROA persisted.

The configuration with 30\% reduction was selected for quantitative comparison against TEER and suture annuloplasty. As summarized in Figure ~\ref{fig:Annuloplasty Repairs} and Table~\ref{tab:annuloplasty-results}, 

In the prolapse model (Table \ref{tab:annuloplasty-results}), both TEER and band annuloplasty restored leaflet coaptation and resulted in similar ROA. Quantitatively, TEER achieved a ROA of 0.106 cm² and a CA of 2.23 cm², while band annuloplasty produced a slightly larger ROA (0.13 cm²) and smaller CA (2.11 cm²). Stress and strain distributions were comparable between the two methods, with both reducing peak leaflet stresses relative to the regurgitant baseline. These findings suggest that for the studied mitral valve, isolated prolapse, TEER may offer slightly superior coaptation performance compared with band annuloplasty.

In the dilated annulus case (Figure \ref{fig:Dilation Repairs}), the differences between techniques were more pronounced. TEER alone failed to achieve complete posterior leaflet capture (Figure \ref{fig:staged capture}, staged capture), leaving a residual ROA of 0.17 cm² and markedly elevated leaflet stresses (95\%ile stress = 2350 kPa). By contrast, band annuloplasty with 30\% annular reduction achieved slightly more effective coaptation, decreasing the ROA to 0.13 cm² while significantly lowering stresses (95\%ile stress = 640.5 kPa) and increasing coaptation area to 2.11 cm². Thus, while TEER was effective in prolapse, band annuloplasty provided a more stable repair in dilation by reducing annular size, improving leaflet contact, and decreasing mechanical loading on the leaflets.

\subsection{Effect of Local Suture Annuloplasty}

\begin{figure}
    \centering
    \includegraphics[width=\textwidth]{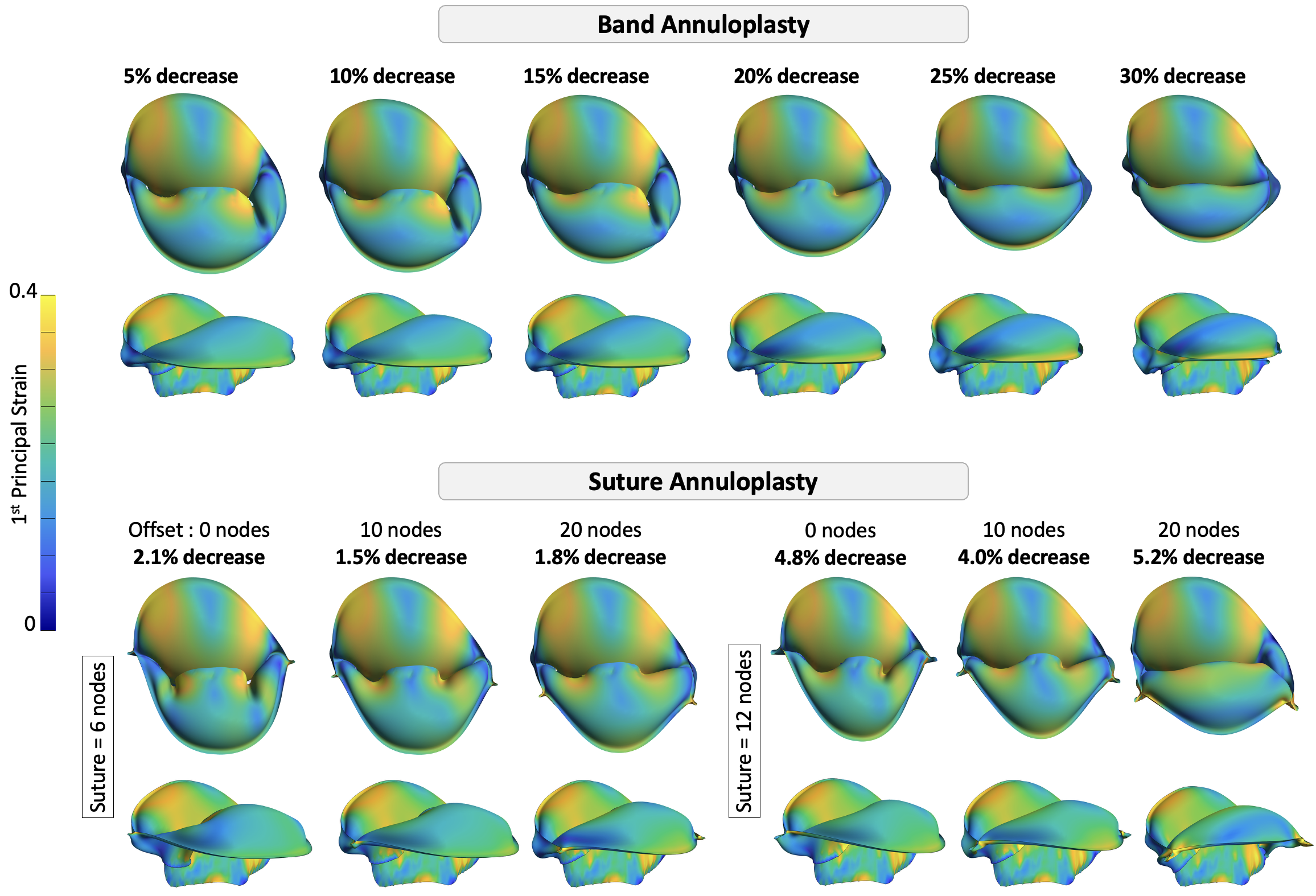}
    \caption{Strain distributions after virtual annuloplasty in the prolapse model. Top: Band annuloplasty with progressive annular circumference reduction ranging from 5\% to 30\%. Bottom: Suture annuloplasty with either 6 or 12 nodes included in suture and varying offsets (0–20 nodes), corresponding to annular circumference reductions of 1.5–5.2\%. Color maps show first principal strain distributions on the valve leaflets after pressurization.}
    \label{fig:Annuloplasty Repairs}
\end{figure}

Next, we investigated an alternative annular reduction technique: suture annuloplasty. This was simulated by virtual sutures at the commissures to reduce the annular circumference. Suture annuloplasty was systematically explored by varying the number (6 vs. 12) and location of sutures by offseting the nodes from the commissures (0 vs. 20 nodes offset), which produced different levels of annular circumference reduction (Figure ~\ref{fig:Annuloplasty Repairs}). Across these simulations, leaflet coaptation generally improved as annular reduction increased, though the magnitude of reduction was modest compared to band annuloplasty. With 12-node sutures and a 20-node offset, corresponding to a 5.2\% annular reduction, complete leaflet closure was achieved. This configuration was selected as representative for quantitative comparison (Table~\ref{tab:annuloplasty-results}). In the prolapse model, the different repair strategies resulted in clearly distinct valve morphologies once the valve was pressurized.

Based on the comparison in Figure \ref{fig:Annuloplasty Repairs}, both annuloplasty techniques produced similar strain distributions on the anterior leaflet, whereas more pronounced differences in strain and morphology were observed on the posterior leaflet following physiological pressurization. Notably, suture annuloplasty achieved a smaller degree of annular reduction compared with band annuloplasty but resulted in more favorable coaptation and mechanical metrics, as shown in Table \ref{tab:annuloplasty-results}. Specifically, suture annuloplasty produced the largest coaptation area ($3.60\,\text{cm}^2$), exceeding both TEER ($2.23\,\text{cm}^2$) and band annuloplasty ($2.11\,\text{cm}^2$), while maintaining comparable regurgitant orifice area (ROA $\approx 0.12\,\text{cm}^2$) and leaflet stress–strain levels. By contrast, band annuloplasty required a greater annular reduction (30\%) yet provided smaller improvements in coaptation.

\begin{table}[h!]
\centering
\begin{minipage}{0.95\linewidth}
    \centering
    \includegraphics[width=\linewidth]{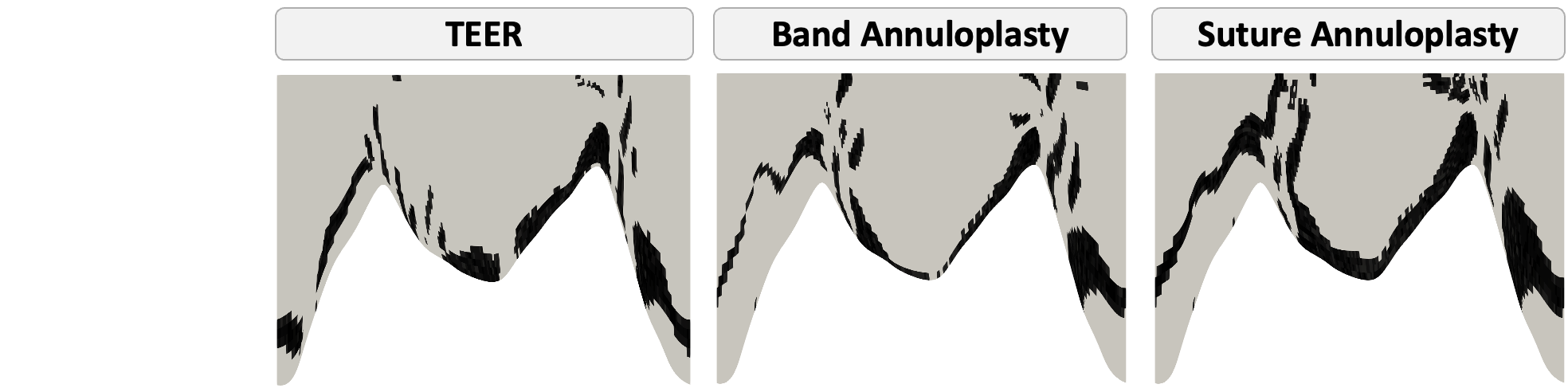} 
\end{minipage}

\vspace{0.5em} 

\begin{minipage}{\linewidth} \centering \begin{tabular}{l>{\centering\arraybackslash}p{3cm}
>{\centering\arraybackslash}p{3cm}
>{\centering\arraybackslash}p{3cm}} \hline & TEER & Band Annuloplasty & Suture Annuloplasty \\ \hline 95\%ile E1 & 0.356 & 0.356 & 0.345 \\ 95\%ile S1 [kPa] & 591.90 & 640.50 & 647.18 \\ ROA [cm$^{2}$] & 0.106 & 0.13 & 0.12 \\ CA [cm$^{2}$] & 2.23 & 2.11 & 3.60 \\ \hline \end{tabular} 
\end{minipage}

\caption{Comparison of TEER, band annuloplasty with 30\% reduction, and suture annuloplasty with 5.2\% reduction. 
Top: leaflet contact regions (black = contact, grey = no contact). 
Bottom: quantitative metrics derived from FE simulations.}
\label{tab:annuloplasty-results}
\end{table}

\subsection{Effect of Combining Annuloplasty with TEER}

\begin{figure}
    \centering
    \includegraphics[width=\textwidth]{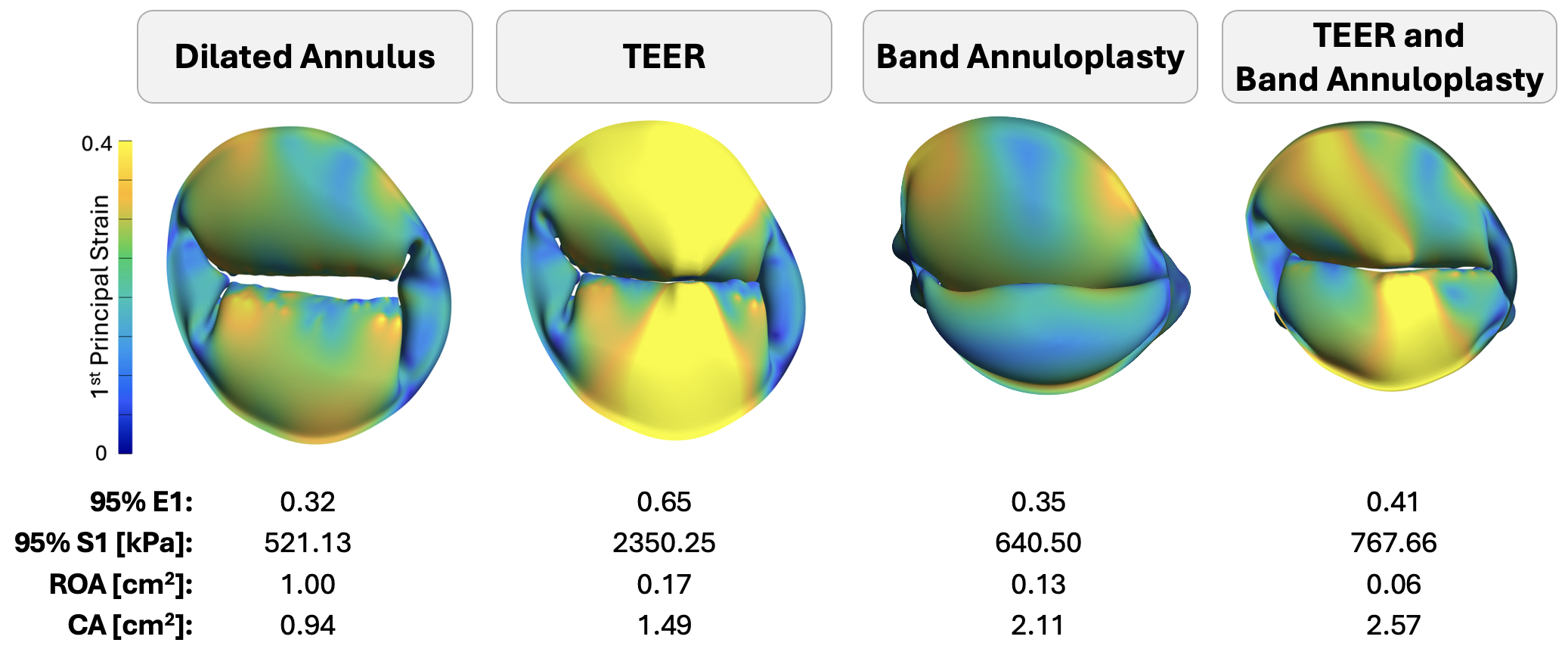}
    \caption{First principal strain distributions in a dilated mitral valve model following different virtual repairs: TEER, band annuloplasty, and combined TEER with band annuloplasty. The TEER simulations used the G4NT clip with staged capture. Band annuloplasty was modeled with a 30\% annular circumference reduction. Quantitative metrics are reported below each case.}
    \label{fig:Dilation Repairs}
\end{figure}

Finally, we combined the TEER clips with annular reduction in a single simulation to assess whether this integrated approach further improves valve mechanics. The outcomes of the dilated annulus case, isolated TEER, isolated band annuloplasty, and the combined repair are presented in Figure ~\ref{fig:Dilation Repairs}. 

In the dilated annulus case, TEER alone reduced ROA from $1.00$ to $0.17$ cm², but induced high leaflet stress (95\%ile $S_1 = 2350.25$ kPa) and strain (95\%ile $E_1 = 0.65$). Band annuloplasty alone (30\% reduction) improved competence with ROA $= 0.13$cm², increased coaptation area (CA= 1.93cm²), and maintained lower leaflet stress ($640.50$ kPa). The combined TEER + band annuloplasty approach provided a favorable balance: ROA was minimized to 0.06 cm², CA was maximized ($2.57$ cm², and leaflet stresses and strains were markedly reduced compared with TEER alone (stress: $2350$ → $767.66$ kPa; strain: $0.65$ → $0.41$). However, this combined approach also produced higher leaflet stresses compared with annuloplasty alone ($767.66$ vs. $640.50$ kPa). Such increases may introduce unfavorable loading conditions, underscoring the need for careful assessment of combined procedures during surgical planning.

\section{Discussion}\label{sec:Discussion}
This study establishes a foundation for simulating TEER and annuloplasty repairs and for comparing valve function and mechanics across different virtual repair strategies using an open-source FEM pipeline. We demonstrated the effect of simulation technique, clip geometry, clip delivery, and variation of annuloplasty techniques on resulting leaflet dynamics and mechanics. Technically, we demonstrated that simulation of "clip release" is essential to determination of realistic valve function. In clinically relevant simulations both functional and mechanical metrics were found to vary across interventional technique and device type, supporting the potential utility of pre-interventional simulation to optimize repair in individual patients in the future.

Previous studies have explored simulation of surgical and transcatheter edge-to-edge repair (TEER) simulations using a variety of modeling strategies. Zhang \textit{et al}.~\cite{zhang2019mechanical} first introduced the concept of FEM-based TEER simulation using “virtual sutures” to approximate the effect of the MitraClip device, though their leaflet model employed a simplified bilayer soft tissue material without anisotropy. Building on this, Kong \textit{et al}.~\cite{kong2020} implemented a more physiologically realistic constitutive law, an anisotropic hyperelastic model based on Holzapfel \textit{et al}.~\cite{HGO_model}, and explicitly modeled clip–leaflet interactions, allowing contact without separation or slip and permitting clip rotation. As the field advanced, Kamakoti \textit{et al}.~\cite{kamakoti2019numerical} performed the first fluid–structure interaction (FSI) simulations, mimicking the clip’s effect on leaflet coaptation without explicitly modeling the device, and demonstrated that clip position strongly influenced leaflet stresses and regurgitation severity. Dabiri \textit{et al}.~\cite{dabiri2022simulation} further expanded FSI analysis to investigate the number and location of clips, highlighting their impact on hemodynamic outcomes. More recently, Rausch\textit{et al}.~\cite{rausch-clip} emphasized the role of annular compliance in FEM simulations, showing that it significantly alters leaflet stresses after virtual clip placement in an abnormal tricuspid valve. 

Building on this body of work, our study extends TEER simulation by presenting multiple transcatheter valve repair techniques—including TEER, band annuloplasty, and suture annuloplasty applied to patient-specific valve geometries. We first explored technical consideration related to clip movement and annular boundary conditions. We introduce a novel “released” clip formulation in FEBio, in which the clip moves together with the valve leaflets based on contact during systole. This approach provides a more physiologically realistic representation of device–tissue interaction compared to previous fixed-clip or suture-based models, and allows for more informative comparisons across TEER and annulus-based repair strategies. In addition, we built upon the work of Rausch\textit{et al}.~\cite{rausch-clip} regarding simulations of annular flexibilty and found that the assumption of a pinned annulus boundary condition was appropriate. We modeled the mitral annulus as "stiffer" than the tricuspid annulus, however validated values of functional mitral and tricuspid properties in functioning hearts are not readily available. Clinically, it has been demonstrated that TEER can have an "annuloplasty effect" due to leaflets pulling in the annulus. As such, this is an area for future exploration both at the tissue level and in functioning heart models incorporating the effect of fluid pressures.

We then explored simulations of different interventional techniques to demonstrate the ability to inform identification of the ideal repair for an individual patient, starting with TEER. Longer clips (G4XT and XTW) generated higher leaflet stress and strain, particularly on the posterior leaflet, while wider clips (G4NTW and XTW) increased leaflet CA, as expected. Importantly, average strain across the entire valve appeared relatively constant across clip types, yet regional analysis revealed increased strain in the posterior leaflet with longer clips. This highlights the importance of evaluating regional rather than global metrics to understand post-clip biomechanics. Clinically, these findings suggest that clip selection may influence long-term leaflet durability and post-repair gradients. Furthermore, when modeling suggests a large leaflet capture gap on either leaflet, our simulations support the use of a “staged leaflet capture” strategy, which increased CA in both prolapse and dilation anatomies.

Our simulations of annulus interventions provided mechanical and clinical metrics, enabling direct comparison between band and suture annuloplasty strategies. For band annuloplasty, $30\%$ reduction was sufficient to achieve full leaflet coaptation, but this would be expected to vary by geometry as previously demonstrated by others\cite{choi2014virtual1, kong2018virtual}. In suture annuloplasty, full closure was achieved by $5.23\%$ reduction in annular circumference. However, location of the suture annuloplasty relative to the commissures influenced the result of the simulation, demonstrating the potential for fine tuning of the suture annuoplasty location to individual patients geometries. For equivalent annular reduction, suture annuloplasty achieved superior coaptation, relative to the band annuloplasty. The strain differences were most pronounced near the annulus, suggesting potential implications for tissue remodeling and long-term valve function.

Notably these methods, with further validation, can be extended to further aspects of pre-intervention decision making, such as clip placement and number of clips, and type and location of transcatheter annuloplasty that may be valuable in complex patient anatomies and especially pediatric populations with atypical and heterogeneous valve morphologies.

\textbf{Limitations and Future Work}
As a foundational study, our simulation pipeline has limitations that create opportunities for future refinement. First, the prolapse and dilation models were derived from a healthy mitral valve by modifying chordal stiffness and annular circumference rather than being reconstructed from patient-specific data. Incorporating real patient-derived geometries will be critical for improving clinical relevance. Second, validation with pre- and post-operative patient data will enhance the predictive accuracy of this framework.

In terms of material modeling, we adopted an isotropic form of the Lee–Sacks constitutive model parameterized for adult mitral valve tissue. While this provides a practical basis, it does not capture the regional heterogeneity of valve leaflets \cite{laurence2019investigation}, which exhibit spatial variations in thickness, stiffness, and anisotropy. Incorporating anisotropy or heterogeneous thickness could complicate comparisons between healthy and diseased states. Another important limitation is the lack of patient-specific material properties. Although Wu et al. \cite{wu2023} demonstrated that variations in leaflet properties can substantially influence simulated valve function, the actual material properties of mitral valve tissue in vivo remain unknown. As such, current models necessarily rely on generalized constitutive descriptions rather than truly patient-specific data. In addition, our simulations did not account for the prestrains that exist in valve tissues \cite{rausch2013mechanics, laurence2022benchtop}. As imaging and data acquisition techniques continue to advance, it may become possible to quantify these prestrains in vivo and incorporate them into future computational models.
Notably, this limitation would be particularly relevant to pediatric or younger valve tissues, which are expected to be more compliant mechanically distinct from adult valves \cite{christierson2025prediction}. Age-dependent constitutive models and microstructural characterization will be critical for improving predictive accuracy in pediatric applications.

For suture annuloplasty, our current implementation applied node-based displacements to achieve annular reduction, making the degree of reduction inherently mesh dependent. In future work, we plan to implement reductions defined in absolute units (e.g., millimeters or centimeters) alongside percentage values, providing surgeons with clearer clinical references and supporting more robust patient-specific surgical planning. 

We also assumed mitral annulus compliance based on prior literature and clinician input due to the absence of reliable in vivo measurements. As imaging and measurement technologies advance, patient-specific annular compliance data can be integrated into the pipeline for improved boundary conditions. Similarly, postoperative imaging data will help refine clip placement strategies, as our simulations showed repair location significantly impacts valve performance. Translationally, this highlights the opportunity to couple simulation-based predictions with intraoperative execution by integrating machine learning–driven real-time motion tracking and delivery guidance.

Additionally, in this study we reported global mechanical metrics using the 95th percentile values across the valve leaflets to compare different repair strategies. While this approach provides a robust overall measure, it does not capture potential regional variations in stress and strain, which may be critical for understanding leaflet function and durability. Future work should therefore incorporate regional analyses to identify localized stress concentrations, particularly in anatomically distinct segments of the mitral valve (e.g., A1–A3, P1–P3) \cite{lichtenberg2020mitral, gao2017modelling}. Implementing such analyses will require careful and reproducible strategies for defining leaflet subregions, which can be especially challenging in pediatric valves due to their greater anatomic heterogeneity and the lack of established leaflet nomenclature.

Finally, we modeled leaflet mechanics without incorporating fluid flow (i.e., no FSI coupling). This structural-only approach allowed us to isolate the mechanical effects of clip deployment and focus on device–tissue interactions, including a realistic representation of clip release. Nonetheless, FSI models provide additional information, particularly on transvalvular hemodynamics, regurgitant volumes, and flow redistribution, which are not captured by structural simulations. Recent work using mitral valve FSI modeling (Razavi et al.\ \cite{razavi2023comparative}) has demonstrated how leaflet deformation and flow fields co-evolve, and comparative studies of structural versus FSI simulations suggest that while technical choices may influence quantitative predictions, qualitative features of leaflet stress distributions and device–tissue mechanics remain consistent \cite{luraghi2018study}. Therefore, we expect that our principal mechanistic conclusions regarding leaflet mechanics and clip–tissue interaction would remain similar under FSI coupling, while quantitative flow-based metrics would be refined and extended. Looking forward, coupling structural modeling with flow dynamics will be an important direction, enabling more comprehensive simulations of post-repair valve performance.

\section{Conclusion}\label{sec:Conclusion}

This study established a foundational FEM open source framework for rapid simulations of transcatheter atrioventriculat valve repair including TEER clip deployment, band annuloplasty, and suture annuloplasty. Our results demonstrate that realistic clip dynamics, device geometry and repair technique distinctly influence leaflet coaptation, stress, and strain. Beyond demonstrating technical feasibility, this work highlights the potential of open-source, image-based FEM frameworks to inform device selection and optimize repair strategies. By providing quantitative clinical metrics and visualizations of valve mechanics, the framework offers a pathway toward patient-specific preoperative planning. With continued refinement through incorporation of patient-derived geometries, age-specific tissue properties, and validation against clinical data, simulations could play a critical role in guiding personalized interventions and improving long-term treatment success. 

\backmatter

\bmhead{Acknowledgments}

Supported by The Cora Topolewski Pediatric Valve Center at CHOP, an Additional Ventures Single Ventricle Research Fund Grant, NIH R01 HL153166, R01 GM083925 (SAM, JAW), T32 HL007915 (DWL, JU), and K25 HL168235 (WW). 

\clearpage
\begin{appendices}

\section{Compliant Annulus}\label{sec:Compliant Annulus Appendix}

\begin{figure}[h!]
    \centering
    \includegraphics[width=\textwidth]{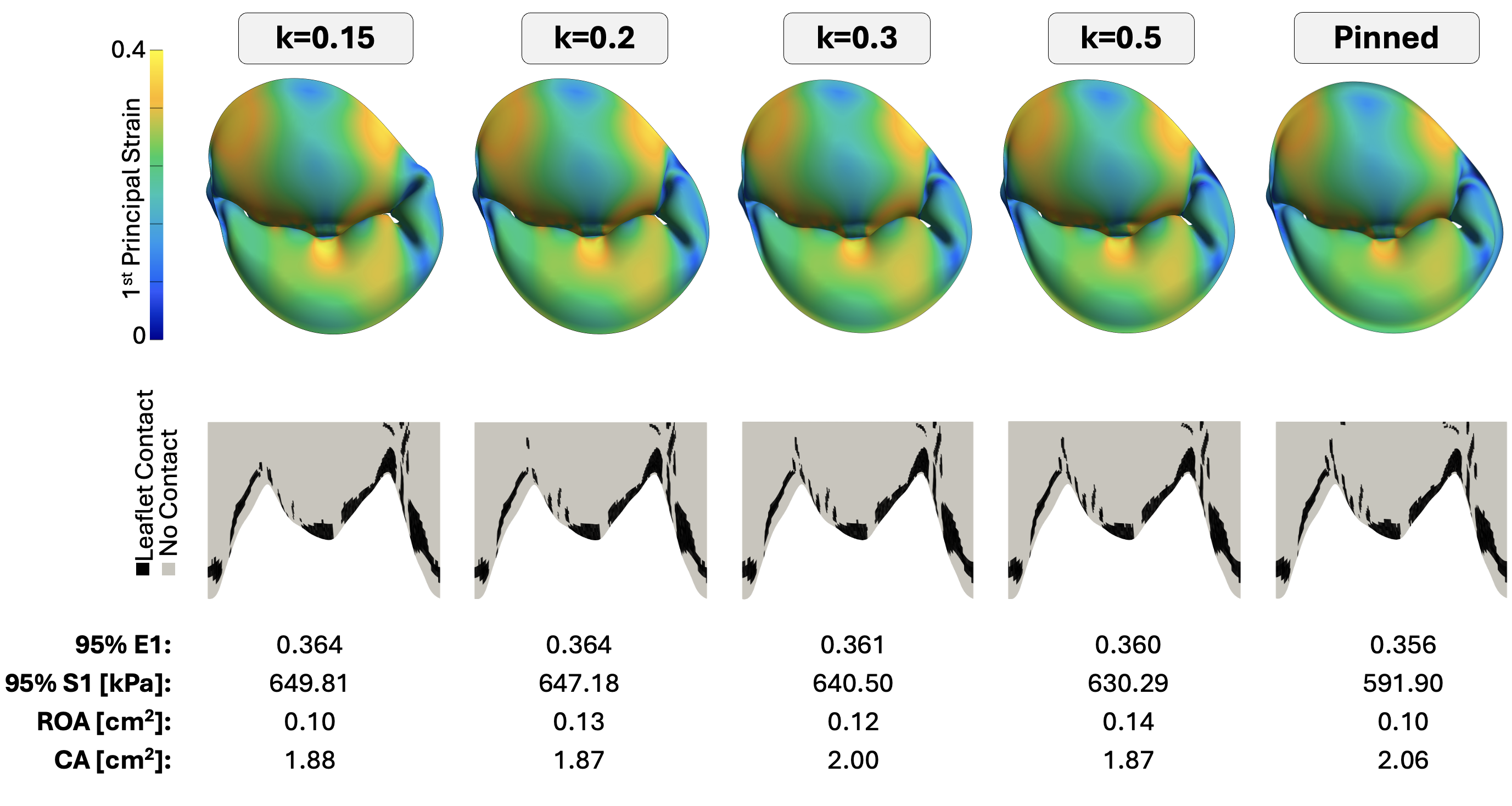}
    \caption{Effect of annular boundary condition stiffness on mitral valve mechanics. Linear springs with different spring constants ($k = 0.15, 0.2, 0.3, 0.5$ N/mm) were applied to annular nodes to represent compliant annulus conditions, and compared against a fully pinned annulus. Top: first principal strain distributions across the leaflets. Middle: leaflet coaptation maps (black = contact, gray = no contact). Bottom: quantitative metrics including 95th percentile strain (E1), 95th percentile stress (S1), regurgitant orifice area (ROA), and leaflet coaptation/contact area (CA).}
    \label{fig:compliant annulus appendix}
\end{figure}

\end{appendices}

\newpage

\end{document}